\documentclass[11pt]{article}

\usepackage{geometry}
\usepackage{amsmath, amsthm, amssymb, amsfonts, upgreek, bm}
\DeclareMathOperator{\argmin}{arg\,min}

\usepackage[numbers,square]{natbib}
\usepackage{setspace}
\usepackage{longtable}
\usepackage{threeparttable}
\usepackage{comment}
\usepackage[utf8]{inputenc}	
\usepackage[T1]{fontenc}	
\usepackage{xcolor}		
\usepackage[colorlinks = true,
            linkcolor = purple,
            urlcolor  = blue,
            citecolor = cyan,
            anchorcolor = black]{hyperref}	
\usepackage{cleveref}
\usepackage{booktabs} 		
\usepackage{nicefrac}		
\usepackage{microtype}		
\usepackage{lineno}		
\usepackage{float}			
\usepackage{subcaption}
\usepackage{algorithm}
\usepackage{color,soul}
\usepackage{algpseudocode}
\usepackage[toc,page]{appendix}

\usepackage{lipsum}		

\usepackage{newfloat}
\DeclareFloatingEnvironment[name={Supplementary Figure}]{suppfigure}
\usepackage{sidecap}
\sidecaptionvpos{figure}{c}

\usepackage{titlesec}
\titlespacing\section{0pt}{12pt plus 3pt minus 3pt}{1pt plus 1pt minus 1pt}
\titlespacing\subsection{0pt}{10pt plus 3pt minus 3pt}{1pt plus 1pt minus 1pt}
\titlespacing\subsubsection{0pt}{8pt plus 3pt minus 3pt}{1pt plus 1pt minus 1pt}

\usepackage{tikz,xcolor,hyperref}
\usetikzlibrary{matrix,positioning,calc}
\usepackage{listofitems} 

\definecolor{lime}{HTML}{A6CE39}
\DeclareRobustCommand{\orcidicon}{
	\begin{tikzpicture}
	\draw[lime, fill=lime] (0,0) 
	circle [radius=0.16] 
	node[white] {{\fontfamily{qag}\selectfont \tiny ID}};
	\draw[white, fill=white] (-0.0625,0.095) 
	circle [radius=0.007];
	\end{tikzpicture}
	\hspace{-2mm}
}
\foreach \x in {A, ..., Z}{\expandafter\xdef\csname orcid\x\endcsname{\noexpand\href{https://orcid.org/\csname orcidauthor\x\endcsname}
			{\noexpand\orcidicon}}
}

  {}              

\title{Crack detection by holomorphic neural networks and transfer-learning-enhanced genetic optimization}

\usepackage{authblk}

\author[1]{Jonas Hund\orcidA{}}
\author[2]{Nicolas Cuenca\orcidB{}}
\author[1\thanks{\tt{Corresponding author: titoan@mpe.au.dk}}]{Tito Andriollo\orcidC{}}

\date{} 

\affil[1]{Department of Mechanical and Production Engineering, Aarhus University, Denmark.}
\affil[2]{SIGMA Clermont, France.}

\begin{document}

\begin{@twocolumnfalse} 
  
\maketitle

\begin{abstract}
A physics-informed machine learning framework based on holomorphic neural networks is introduced for detecting cracks in two-dimensional solids from strain or displacement data. Crack detection is formulated as an inverse problem in which the crack size, orientation, and location are treated as unknowns. The problem is solved using genetic optimization, where the fitness function is evaluated by expressing the solution of the corresponding plane elasticity problem in terms of holomorphic potentials, which are then determined through the training of two holomorphic neural networks. As the potentials satisfy equilibrium and traction-free conditions along the crack faces a priori, the training proceeds quickly based solely on boundary information. Training efficiency is further improved by splitting the genetic search into long-range and short-range stages, enabling the use of transfer learning in the latter. The new strategy is tested on three benchmark problems, showing that an optimal number of training epochs exists that provides the best overall performance. A comparison is also made with a popular crack detection approach that uses XFEM to compute the model response. Under the assumption of identical stress-field representation accuracy, the proposed method is found to be between 7 and 23 times faster than the XFEM-based approach. Furthermore, the proposed method appears to be less sensitive to noise in the input data. Overall, the present findings demonstrate that combining genetic optimization with holomorphic neural networks and transfer learning offers a promising avenue for developing crack detection strategies with higher efficiency than those currently available.
\end{abstract}

\vspace{0.20cm}

\textbf{Keywords}: linear elastic fracture mechanics, physics-informed neural networks, holomorphic neural networks, crack detection, structural health monitoring

\vspace{0.20cm}

\end{@twocolumnfalse}
\onehalfspacing

\begin{longtable}{|l l|}
\hline
\textbf{Nomenclature} & \\ & \\
\textbf{Abbreviations} & \\
GA & genetic algorithm \\
HNN & holomorphic neural network \\
KM & Kolosov-Muskhelishvili \\ 
PINN & physics-informed neural network \\ 
TL & transfer learning \\
XFEM & extended finite element method \\ & \\
\textbf{Symbols} & \\
$a, \alpha $ & crack half length and orientation angle \\
$ E $ & crack detection error \\
$H$ & space of holomorphic functions \\
$i$ & imaginary unit \\
$L$ & plate half side length \\
$\mathcal{L}$ & loss function \\
$ m $ & noise level \\
$\bm{n}$ & normal outward unitary vector with respect to a curve \\
$N_I$ & initial crack population size \\
$N_R$ & number of cracks removed at each generation \\
$N_O$ & number of crack offspring at each generation \\ 
$N_S$ & number of cracks passed on to the short-range search algorithm \\
$N_D$ & number of crack duplicates in the short-range search algorithm \\
$N_\Delta$ & crack population decrease after each generation \\ 
$N_u,N_\sigma$ & number of training points on $\Gamma_u,\Gamma_\sigma$ \\ 
$\mathbb{R},\mathbb{C}$ & real and complex field \\
$\bm{t_0}, \bm{u_0}$ & prescribed traction and displacement on the boundary \\ 
$\bm{u}$ & Displacement vector \\
$\bm{p}$ & Vector with crack geometrical parameters \\
$x,y$ & 2D space coordinates \\
$z$ & complex space coordinate \\
$\Gamma_u,\Gamma_\sigma$ & Dirichlet and Neumann portions of the boundary \\
$\bm{\upvarepsilon}$ &  strain tensor \\
$\kappa$ & Kolosov's constant \\
$\lambda, \mu$ & Lamé first and second parameter \\
$\bm{\upsigma}$ &  Cauchy stress tensor \\
$\varphi, \psi, \omega$ & Kolosov-Muskhelishvili potentials \\
$\chi, \gamma $ & auxiliary holomorphic functions \\
$\Omega, \partial\Omega$ & solid body domain and associated boundary \\
$\mathcal{P}$ & search space \\
$P_M, \delta_M$ & mutation probability and magnitude \\
$\mathcal{F}_t, \mathcal{F}_s$ & fitness thresholds for termination and short-range search \\
$ Y $ & uniformly distributed random variable \\
$\overline{\cdot}$ & conjugation operator \\
$\check{\cdot}$ & double conjugation operator \\
\hline
\end{longtable}
\addtocounter{table}{-1}

\section{Introduction}\label{sec:introduction}

The capability to detect, locate, and evaluate damage or degradation in structures to ensure their reliability and performance throughout their service life is crucial for modern society. In this context, research on structural health monitoring has received considerable attention over the past decades \cite{gharehbaghi_critical_2022}. Structural health monitoring can be defined as the assessment of a structure’s condition through integrated sensing, data acquisition, and analysis techniques \cite{worden_fundamental_2007,charles_r_farrar_structural_2012}. Several well-established sensing methodologies are now available to detect early signs of degradation, predict structural failures, and ensure overall safety. These include temperature sensing, electrical and electrochemical sensing, local strain measurement, ultrasonic testing, automated visual inspection, vibration-based methods, and acoustic emission techniques \cite{mardanshahi_sensing_2025}. Based on the recorded signals, damage detection is typically conducted either using a reference model of the structure (physics-based approaches) or through signal-processing-based techniques (data-driven approaches) \cite{gharehbaghi_critical_2022}.

In physics-based approaches, damage is detected via an inverse procedure that aims to minimize the discrepancies between the model response and the sensor data \cite{prashant_m_pawar_structural_2011}. In the specific case of crack detection based on strain measurements, this procedure is formulated as an optimization problem in which model parameters such as crack location, orientation, and size are progressively updated until the model response matches the sensor data. The combination of the extended finite element method (XFEM) and genetic algorithms (GAs) has proved particularly effective for solving such inverse problem \cite{rabinovich_xfem-based_2007,waisman_detection_2010,chatzi_experimental_2011,agathos_multiple_2018}. This effectiveness arises from the ability of XFEM to evaluate the structural response for different crack parameter values without the need for remeshing, together with the capacity of GAs to explore a complex solution space without relying on gradient information, thereby reducing the risk of being trapped in local minima.

While the GA-XFEM combination offers excellent flexibility and robustness, its computational cost remains a significant drawback. Consequently, several strategies have been investigated to accelerate the inverse problem solution, including surrogate modeling approaches based on model-order reduction techniques \cite{kerfriden_localglobal_2012,benaissa_crack_2016} and, more recently, machine learning methods \cite{yoon_deep_2022,huang_deep_2023,thananjayan_scaled_2024}. Although these approaches provide advantages in specific scenarios, the former are limited by the need to update the reduced space on-the-fly, whereas the latter require large amounts of training data, which may not be available. 

In an effort to reduce the reliance on experimental training data, several studies have proposed hybrid approaches where machine learning is supported by computational mechanics tools \cite{daghigh_review_2024}. For instance, \citet{lee_structural_2023} proposed a hybrid framework in which deep neural networks are trained on damage scenarios generated from a finite element model updated using experimental modal data. \citet{zhong_structural_2025} introduced a phase-space matrix representation of vibration signals combined with convolutional neural networks, achieving high robustness to measurement noise and finite element modeling uncertainty, even with a limited number of sensors. For composite structures, \citet{nishioka_development_2025} developed a defect localization strategy that couples finite element simulated stress fields with graph neural networks, exploiting mesh topology through graph attention mechanisms to predict three-dimensional delamination cracks. In large-scale marine structures, \citet{bardiani_real-time_2026} integrated finite element modeling with machine learning to enable real-time damage detection and localization under extreme operational conditions, illustrating the scalability of hybrid machine learning–computational mechanics systems for complex geometries.

On a different track, the extensive amount of data required to train machine learning models has also prompted numerous investigations in the rapidly growing field of physics-informed machine learning \cite{karniadakis2021physics,toscano_pinns_2025}, and in particular, physics-informed neural networks (PINNs) \cite{lagaris1998artificial, raissi2019physics}. These approaches require substantially less data, as they rely on enforcing physical constraints. In the simplest form of the PINN approach, a neural network is employed as a global ansatz function to solve a given boundary value problem. The network parameters (weights and biases) are optimized during training by minimizing a loss function that includes the residuals of the governing partial differential equations, the boundary conditions, and any deviations from available experimental data.

While a large number of studies have applied PINNs and their variants to solid mechanics, approaches tailored to problems within the scope of linear elastic fracture mechanics remain relatively less explored \cite{goswami_physics-informed_2022,manav_phase-field_2024, zheng_physics-informed_2022,Baek_2022,gu2023,chen2024crack,leon_exploring_2025}. One possible reason is the discontinuous nature of the solution across the crack surfaces, combined with the singular behavior near the crack tips, which necessitates the adoption of specialized strategies to mitigate convergence and accuracy issues. On the other hand, as noted by \citet{zhang_analyses_2022}, the meshless nature of PINNs makes them potentially ideal for solving inverse problems involving the identification of material defects based on strain or displacement measurements at discrete sensor locations. Furthermore, transfer learning (TL) can significantly enhance the efficiency of PINNs when solving sequences of problems that differ only slightly in domain geometry, as demonstrated by \citet{chen2024crack} for crack propagation simulations.

In addition to these developments, the emergence of holomorphic neural networks (HNNs) has significantly accelerated training for problems whose solutions can be represented by holomorphic functions \cite{calafa2024}. A notable example is plane linear elasticity, where the Kolosov–Muskhelishvili (KM) representation \cite{muskhelishvili1977} allows expressing the solution in terms of two complex-valued, holomorphic potentials. Accordingly, HNNs have been shown to accelerate the solution of plane elasticity problems by one to two orders of magnitude compared to standard, real-valued PINNs \cite{calafa2024}. This improvement arises because the holomorphic formulation focuses solely on learning the complex KM potentials that satisfy the boundary conditions, while the governing equations are inherently satisfied a priori.

Recently, two enrichment strategies have been proposed to further improve the accuracy and efficiency of HNNs for fracture mechanics problems \cite{calafa_solving_2025}. The first involves enriching the HNNs with the square-root term from Williams’ series, while the second leverages Rice’s global representation of the solution for a straight crack to decouple the holomorphic part of the solution from the singular, non-holomorphic terms. In addition to significantly improving accuracy in the presence of cracks, both strategies were shown to be highly compatible with TL, enabling a ten-fold reduction in training time for repeated computations. Remarkably, preliminary comparisons indicated that, for comparable levels of target accuracy, the computational efficiency of enriched HNNs when using TL was nearly comparable to that of Abaqus XFEM \cite{calafa_solving_2025}.

In light of the above, the present work proposes a novel and efficient framework for crack detection in two-dimensional linear elastic solids that leverages HNNs and transfer learning. This framework combines the enriched HNN formulation introduced by \citet{calafa_solving_2025} with a modified genetic optimization algorithm in which the crossover and mutation search stages are decoupled. The remainder of this article is organized as follows. Section \ref{sec:theory} provides the necessary theoretical background, including the complex-valued formulation of plane linear elasticity, the corresponding complex representation of the solution for an internal crack, and an introduction to HNNs. Section \ref{sec:inverse_problem} defines the crack detection problem and describes two solution strategies: the state-of-the-art approach based on GA-XFEM, and the new approach combining HNNs with GA. The performance of the new approach is thoroughly analyzed in Section \ref{sec:results} in terms of convergence, noise sensitivity, and efficiency in relation to the standard GA-XFEM approach. Finally, Section \ref{sec:conclusions} summarizes the main findings and conclusions.

\section{Theory}\label{sec:theory}
\subsection{Complex representation of plane linear elasticity} \label{sec:complex_repr_lin_elasticity}
The equations of linear elasticity for an isotropic and homogeneous solid occupying a region $\Omega\subset \mathbb{R}^2$ can be written as
\begin{equation}
\begin{cases}
    \displaystyle \nabla \cdot \bm{\upsigma} = \bm{0},\\ 
    \displaystyle \bm{\upsigma} = 2\mu \bm{\upvarepsilon} + \Tilde{\lambda} \text{Tr}(\bm{\upvarepsilon}) \mathbf{I},\\ 
    \displaystyle \bm{\upvarepsilon} = \frac{\nabla \bm{u} + \nabla \bm{u}^T}{2},
    \label{eq:linearelasticity}
\end{cases}
\end{equation}
where $\bm{\upsigma}$ denotes the Cauchy stress tensor, $\bm{\upvarepsilon}$ is the infinitesimal strain tensor and $\bm{u}$ is the displacement vector. Furthermore, $\lambda,\mu\in\mathbb{R}$ are the Lamé first and second parameters, respectively, $\text{Tr}(\cdot)$ denotes the trace operator, $\mathbf{I}$ is the identity tensor and $\Tilde{\lambda} = \lambda $ for plane strain, whereas $\Tilde{\lambda} = \left(2 \lambda \mu\right)/\left(\lambda + 2 \mu \right) $ for plane stress.

Traction and displacement boundary conditions are assumed to be applied on the solid boundary $\partial\Omega$:
\begin{equation}
\label{eq:BC}
    \begin{cases}
        \bm{\sigma} \cdot \bm{n} = \bm{t_0}, &\hspace{5mm} \text{ on } \Gamma_\sigma, \\
    \bm{u} = \bm{u}_0, &\hspace{5mm} \text{ on } \Gamma_u, 
    \end{cases}
\end{equation}
where $\bm{t_0}$ is the imposed traction and $\bm{u}_0$ the prescribed displacement. In order to guarantee well-posedness, we assume that $\overline{\partial\Omega} = \overline{\Gamma_\sigma} \cup \overline{\Gamma_u} $ and $\Gamma_\sigma\cap \Gamma_u = \emptyset$.

The boundary value problem defined by \Cref{eq:linearelasticity,eq:BC} can be equivalently formulated in terms of complex potentials by leveraging the Kolosov-Muskhelishvili (KM) representation \cite{kolosov1909application,muskhelishvili1977}. Let us define $H(\Omega)$ the set of holomorphic functions on $\{z=x+iy:(x,y)\in\Omega\}$. Then, assuming the domain $\Omega$ to be simply-connected, $\bm{\upsigma}$ and $\bm{u}$ satisfy \Cref{eq:linearelasticity} if and only if there exist $\varphi,\psi\in H(\Omega)$, termed the KM potentials, such that
\begin{equation}\label{eq:km}
    \begin{cases}
    \sigma_{xx} = \text{Re}\left(2\varphi' - \overline{z}\varphi''-\psi'\right), \\
    \sigma_{yy} = \text{Re}\left(2\varphi' + \overline{z}\varphi''+\psi'\right), \\
    \sigma_{xy} = \text{Im}\left(\overline{z}\varphi''+\psi'\right), \\
    u_x = \frac{1}{2\mu}\text{Re}\left(\kappa \varphi - z \overline{\varphi'} - \overline{\psi}\right), \\ 
    u_y = \frac{1}{2\mu}\text{Im}\left(\kappa \varphi - z \overline{\varphi'} - \overline{\psi}\right), \\ 
    \end{cases}
\end{equation}
where $\kappa:= (\tilde{\lambda} + 3\mu)/(\tilde{\lambda} +\mu)$ is the Kolosov constant, $\overline{(\cdot)}$ denotes the complex conjugate, and $(\cdot)'$ denotes the complex derivative.

While $\varphi$ and $\psi$ represent the standard choice of complex potentials, for problems involving cracks it can be convenient to work with an alternative potential defined as
\begin{equation}\label{eq:omega}
    \omega(z):= \psi(z) + z \varphi'(z).
\end{equation}
As the above definition implies that
\begin{equation}\label{eq:psi}
    \psi(z) = \omega(z) - z \varphi'(z).
\end{equation}
the solution to a linear elastic problem can be equivalently expressed either in terms of $(\varphi, \psi) $ or in terms of $(\varphi,\omega)$.

\subsection{Solid with internal crack} \label{sec:solution_internal_crack}
\begin{figure}[!ht]
        \centering
        \includegraphics[width=0.5\textwidth]{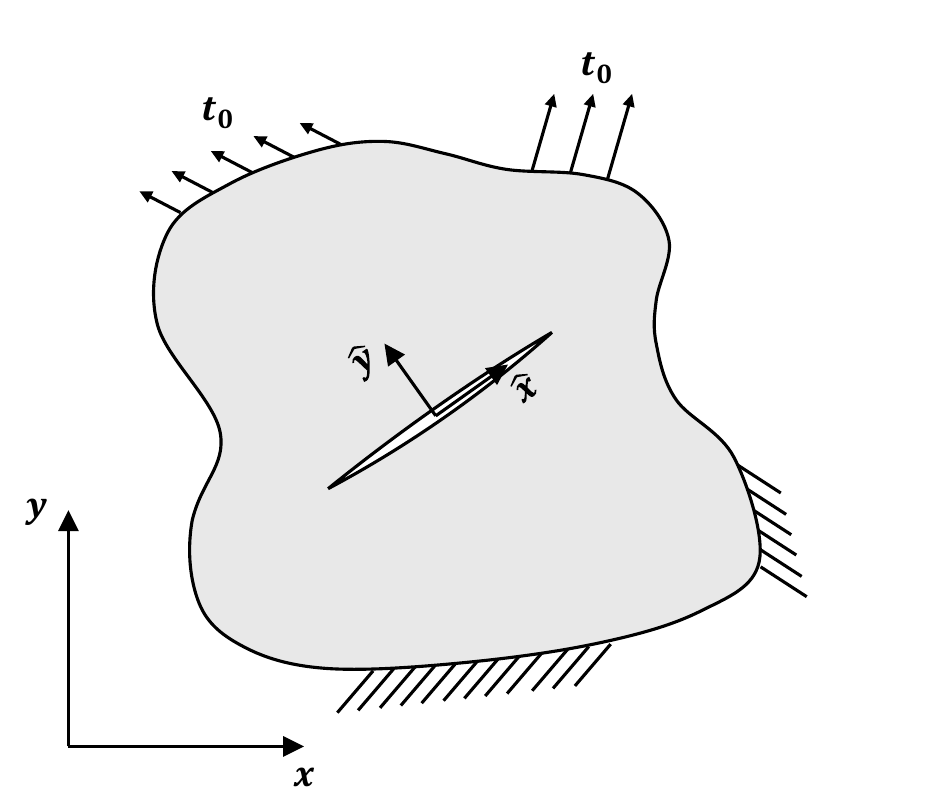}
        \caption{Linear elastic solid containing a straight internal crack.}
        \label{fig:body_with_crack}
\end{figure}

Let us consider the special case of a solid containing a straight internal crack of length $2a$ with traction-free faces, see \Cref{fig:body_with_crack}. 
If a local coordinate system $(\hat{x}, \hat{y})$ is defined such that the origin is located at the center of the crack and the crack lies on the $\hat{x}$-axis, then the complex potentials can be expressed as \cite{woo1989,wang2003boundary}
\begin{equation}\label{eq:rice_mod}
\begin{cases}
    \hat{\varphi}(\hat{z}) = \chi(\hat{z}) \sqrt{\hat{z}-a}\sqrt{\hat{z}+a} + \gamma(\hat{z}),\\
    \hat{\omega}(\hat{z}) = \check{\chi}(\hat{z}) \sqrt{\hat{z}-a}\sqrt{\hat{z}+a} - \check{\gamma}(\hat{z}),
\end{cases}
\end{equation}
where $\chi(\hat{z})$ and $\gamma(\hat{z})$ are two auxiliary functions which are holomorphic everywhere, $\sqrt{\cdot}$ denotes the principal branch of the complex square root, and the operator $\check{\cdot}$ is the double conjugation of a complex function defined by $ \check{g}(z) := \overline{g(\overline{z})}$. A key feature of the representation \eqref{eq:rice_mod} is that any choice of $\chi,\gamma\in H(\Omega)$ generates an elastic solution that automatically fulfills the boundary conditions along the crack faces. That is, an elastic solution featuring a displacement jump across the crack faces and square root stress singularity at the crack tips is produced from two functions $\chi$ and $\gamma$ that, by contrast, are continuous and infinitely smooth everywhere, also on the crack. This property offers a significant advantage for the application of HNNs, as discussed in the next section.

The potentials $\hat{\varphi}, \hat{\omega} $ in \Cref{eq:rice_mod} are denoted with a hat to emphasize that they are valid in the local coordinate system. The standard potentials $\varphi, \psi $ in the global coordinate system $(x, y)$ can be obtained by applying \Cref{eq:psi} and the roto-translation formulae reported in \citet{calafa_solving_2025}. Assuming that the $\hat{x}$-axis is rotated anti-clockwise by an angle $\alpha $ with respect to the x-axis and the crack center is located at $z_c\in \mathbb{C}$, the resulting expressions are 
\begin{equation}\label{eq:rototraslatedpotentials}
    \begin{cases}
        \varphi(z) = e^{i\alpha}\hat{\varphi}(\hat{z}(z)), \\
        \psi(z) = e^{-i\alpha}\hat{\omega}(\hat{z}(z)) - \left( e^{-i\alpha}\hat{z}(z) + \overline{z_c} \right) \hat{\varphi}'(\hat{z}(z)),
    \end{cases}
\end{equation}
where
\begin{equation}\label{eq:z_hat}
        \hat{z}(z) = e^{-i\alpha}(z-z_c)
\end{equation}
From knowledge of $\varphi, \psi $, stress and displacement components in the global coordinate system can be computed via the KM formulae \eqref{eq:km}.

\subsection{Solution using holomorphic neural networks}\label{sec:HNNs}
A holomorphic neural network (HNN) is a complex-valued artificial neural network employing holomorphic activation functions. As demonstrated in \cite{calafa2024}, the choice of holomorphic activation functions – typically the complex exponential – has two important consequences. First, the network output remains holomorphic across the entire complex plane for any values of the network parameters. Second, the network can approximate any holomorphic function to an arbitrary degree of accuracy, provided a sufficient number of layers and neurons is used. These properties make HNNs particularly suitable for solving boundary value problems in which the solution can be expressed in terms of holomorphic potentials \cite{calafa2024}. In fact, by leveraging the principles of the PINN approach, HNNs can effectively learn such potentials by minimizing a loss function that quantifies the violation of the boundary conditions. In contrast to the standard PINN approach, no explicit enforcement of the governing differential equations is required, as they are inherently satisfied. This leads to significant gains in both accuracy and computational efficiency \cite{calafa2024, calafa_solving_2025,calafà2026}.

For the internal crack problem of \Cref{fig:body_with_crack}, HNNs can be used to compute the solution as follows \cite{calafa_solving_2025}. First, a set of training points $\{z_i\}$ is generated along the domain boundary by sampling from a uniform distribution. Next, each training point is passed as input to a pair of HNNs, which provide the values of the two auxiliary functions $ \chi,\gamma $ as output. From these, the standard potentials $\varphi, \psi $ are calculated via \Cref{eq:rice_mod,eq:rototraslatedpotentials} and subsequently used to find stress and displacement components through automatic differentiation in combination with the KM formulae \eqref{eq:km}. The loss function is then computed as
\begin{equation}\label{eq:loss_simple}
    \mathcal{L}:=s_{u}\mathcal{L}_{u} + s_{\sigma}\mathcal{L}_{\sigma} ,
\end{equation}
 where $s_{u}, s_{\sigma}$ are scalar hyperparameters and the terms $\mathcal{L}_{u}$ and $\mathcal{L}_{\sigma}$ quantify the extent to which the boundary conditions \eqref{eq:BC} are violated. They are defined as 
\begin{align}
\mathcal{L}_{u}&:= \frac{1}{N_u}\sum_{z_i \in \Gamma_u} \|\bm{u}(z_i) - \bm{u}_0(z_i)\|_2^2, \\
\mathcal{L}_{\sigma}&:= \frac{1}{N_\sigma}\sum_{z_i \in \Gamma_\sigma} \|\bm{\upsigma}(z_i) \cdot \bm{n} - \bm{t}_0(z_i)\|_2^2
\end{align}
where $N_u$ and $N_\sigma$ denote the number of training points along $\Gamma_u$ and $\Gamma_\sigma$, respectively. Finally, the loss function gradient is computed through backpropagation and used to update the parameters of the two HNNs. The process is repeated until the desired level of convergence is attained, with each repetition cycle defining a training epoch.

It is worth noting that the use of the representation \eqref{eq:rice_mod} brings two important advantages. First, in contrast to the standard KM potentials, the two functions $ \chi,\gamma $ are holomorphic everywhere, meaning that they are ideally suited to be approximated using HNNs. Second, there is no need to place training points on the crack faces, since the corresponding boundary conditions are satisfied exactly by construction.

\section{Crack detection problem}\label{sec:inverse_problem}
\subsection{Formulation}
Inverse problems involve identifying a set of unknown model parameters from knowledge of the expected model response. For the problem of crack detection based on strain sensor data, the unknown parameters are geometrical quantities describing the crack’s shape, size, and location, while the response is represented by strain measurements. To leverage the theory presented in \Cref{sec:theory}, the present work focuses on the specific case of detecting an internal crack in a two-dimensional linear elastic solid. It is assumed a priori that the crack is straight and traction-free, whereas its size, orientation, and location are unknown and must be determined from strain data collected at a discrete set of locations. 

Following previous studies \cite{waisman_detection_2010,agathos_multiple_2018}, the problem is formulated as follows:
\begin{equation}\label{eq:optimization_equation}
    \bm{p}^* = \underset{\bm{p} \in \mathcal{P}} \argmin \, \mathcal{F}(\bm{p}) 
\end{equation}
where $\bm{p}$ is a collection of geometrical parameters describing the crack, $\mathcal{P}$ is the set of admissible values of the former parameters, and $\mathcal{F}$ is the objective function to be minimized. The latter is defined as
\begin{equation}\label{eq:obj_function}
    \mathcal{F}(\bm{p}) =  \frac{\sum_{n=1}^{n_s} \Vert \bm{\upvarepsilon}(\bm{p}, z_n) - \bm{\upvarepsilon}_n \Vert_2^2}
    { \sum_{n=1}^{n_s} \Vert \bm{\upvarepsilon}_n \Vert_2^2} 
\end{equation}
where $\bm{\upvarepsilon}(\bm{p}, z_n)$ is the strain predicted by the model at the $z_n$ location of the n-th sensor, $\bm{\upvarepsilon}_n$ is the strain measured by the n-th sensor, and $n_s$ is the total number of sensors. Here, and in the remainder of the article, the term "model" refers to the mathematical representation of the solid defined by its geometry and the linear elastic relationships \labelcref{eq:linearelasticity,eq:BC} that specify the mechanical behavior.   

\subsection{XFEM-based genetic optimization}\label{sec:GA-XFEM}
As anticipated in the introduction, the optimization problem defined by \Cref{eq:optimization_equation} is most commonly solved using GAs. The fundamental idea behind genetic optimization is to mimic the biological processes of natural evolution and the survival of the fittest individuals \cite{Book_Michalewicz, prashant_m_pawar_structural_2011}. By analogy, the set of candidate solutions represents a population of individuals, where each individual is encoded by a set of genes. A random generation of such gene sets provides the initial population (\textit{initialization}). During each iteration of the algorithm – referred to as \textit{generation} –, offspring are produced from the fittest individuals of the previous generation (\textit{selection}). The fitness of each individual is evaluated using the objective function associated with the optimization problem. Reproduction occurs through recombination of genes (\textit{crossover}) and random gene variations (\textit{mutation}), both applied with a certain probability. The goal of this evolutionary process is to promote the survival of the fittest individual, which produces the largest number of offspring and ultimately dominates the population over successive generations.

\begin{figure}[!ht]
        \centering
        \includegraphics[width=0.8\textwidth]{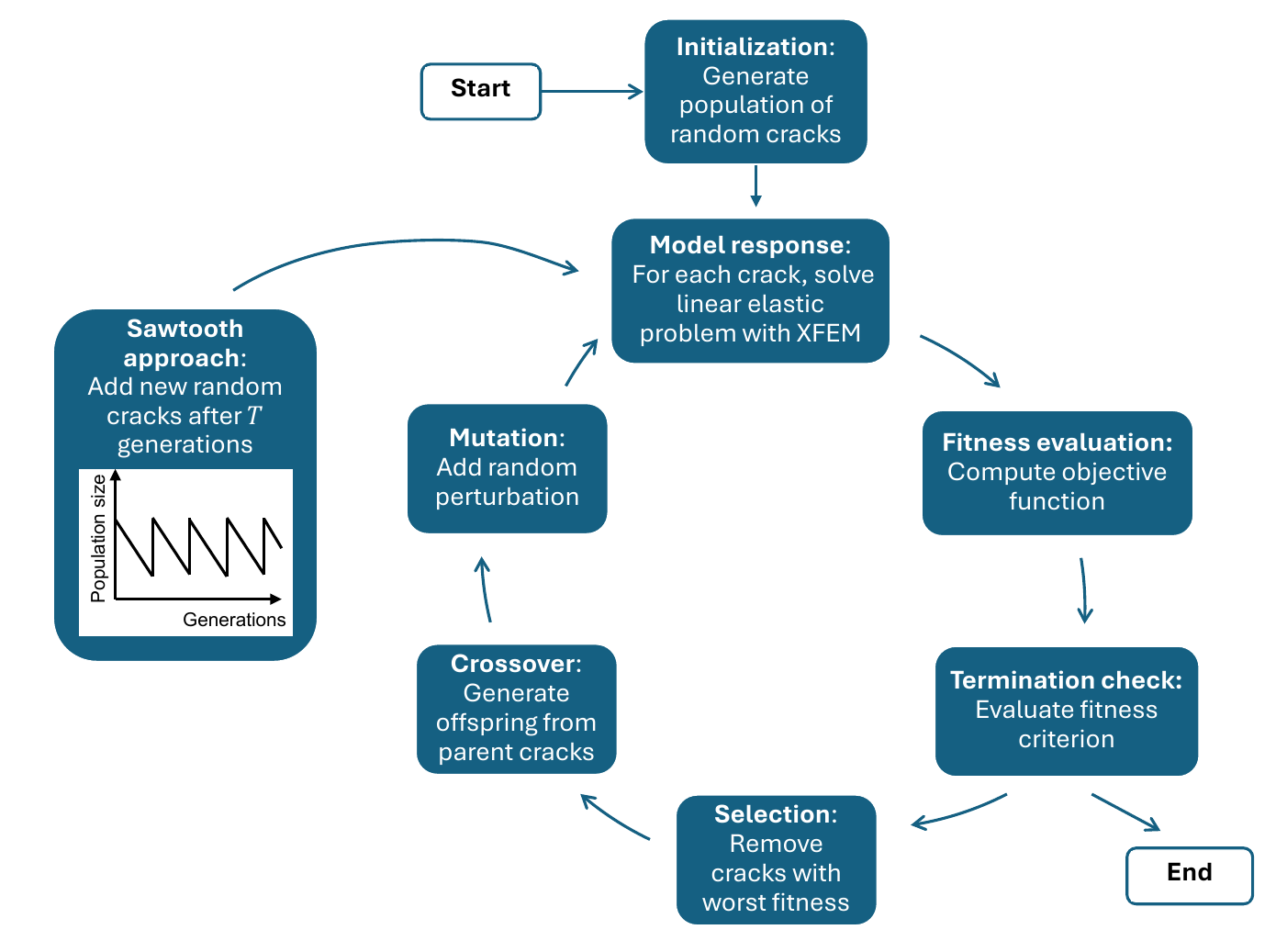}
        \caption{Reference crack detection algorithm proposed by \citet{waisman_detection_2010}, which combines genetic optimization with XFEM.}
        \label{fig:waisman_algorithm}
\end{figure}

Several variations of GAs have been proposed in the literature, most of which can be applied to the crack detection problem defined by \Cref{eq:optimization_equation}. In this work, the objective is to adopt a well-known and robust implementation that can serve as a reference for the benchmarks presented in \Cref{sec:num_experiments}. Accordingly, an algorithm closely matching the widely used approach proposed by \citet{waisman_detection_2010} is employed, whose schematic representation is shown in \Cref{fig:waisman_algorithm}. 

The algorithm assumes that the geometrical parameters describing the crack are the crack tip coordinates $z^+,z^- \in \mathbb{C} $, i.e. $\bm{p} = \{z^+,z^-\} $. From these, the crack half length, center location and orientation can be computed as
\begin{equation}
    a = \frac{|z^+ - z^-|}{2}, \quad
    z_c = \frac{z^+ + z^-}{2}, \quad
    \alpha = \arg (z^+ - z^-)
\end{equation}
The algorithm first step is to create an initial population of $N_{I}$ random cracks subjected to the constraint that the crack tips must belong to the convex search space $ \mathcal{P} \subset \Omega $. Then, for each crack, the model response at the sensor locations is computed using Abaqus XFEM and used to calculate the crack fitness, quantified by the objective function of \Cref{eq:obj_function}. If at least one crack is found for which 
\begin{equation}\label{eq:fitness_threshold}
    \mathcal{F}(\bm{p}) < \mathcal{F}_{t},
\end{equation}
where $\mathcal{F}_{t}$ is a selected threshold, or if the maximum number of generations is exceeded, the algorithm terminates. Otherwise, the
$N_R $ cracks with the worst fitness are removed from the population, while the other $N_P$ cracks undergo arithmetical crossover \cite{Book_Michalewicz}. This operation consists in generating $N_{O}$ crack offspring, where each child crack is generated from two randomly selected parent cracks. Specifically, 
\begin{equation}\label{eq:crossover}
\begin{split}
    &(z^+)_O = \beta(z^+)_{P1} + (1-\beta)(z^+)_{P2} \\
    &(z^-)_O = \beta(z^-)_{P1} + (1-\beta)(z^-)_{P2}
\end{split}
\end{equation}
where the subscripts $ (\cdot)_O, (\cdot)_{P1}, (\cdot)_{P2}$ refer to the offspring, first parent and second parent crack, respectively, and $\beta \in [0,1]$ is a random variable with uniform distribution. After crossover, mutation is applied to each crack with a probability $ P_M $, introducing a perturbation to the crack tip coordinates defined as
\begin{equation}\label{eq:mutation}
\begin{split}
    &(z^+)_M = z^+ + \delta_1 + i\delta_2 \\
    &(z^-)_M = z^- + \delta_1 + i\delta_2
\end{split}
\end{equation}
where the subscript $ (\cdot)_{M}$ refers to the values after mutation, $\delta_1, \delta_2 \in [-\delta_M,+\delta_M] $ are uniformly distributed random variables and the parameter $\delta_M \in \mathbb{R}$ controls the mutation magnitude. Finally, the model response is evaluated for the updated crack population and the previous steps are repeated. 

It is worth noting that since the difference $N_R-N_O:=N_{\Delta} $ is normally set to be greater than zero, the crack population size decreases at each generation. To compensate for this, new randomly generated cracks are added to the population every $T>1$ generations. This leads to a population size that varies with the number of generations according to a sawtooth function. As discussed by \citet{waisman_detection_2010}, this partial re-initialization of the population is beneficial for convergence, as it enhances the algorithm ability to move away from local minima.

\subsection{HNN-based genetic optimization enhanced via transfer learning}\label{sec:GA-HNNs}
The GA presented in the previous section employs the fundamental operations of crossover and mutation to explore the search space. Crossover performs a long-range search by generating crack offspring that may differ substantially from the parent cracks in terms of position of their tips. By contrast, mutation performs a short-range search by introducing small variations in the crack tip coordinates. This distinction is crucial when considering the replacement of XFEM with HNNs for evaluating the model response. 

The primary motivation for adopting HNNs is to leverage TL, which, in general terms, involves using the network parameters obtained from one problem as the initial condition for solving a subsequent, related problem. TL is effective when the solutions to the two problems are sufficiently similar. In the present context, this implies that TL is beneficial only when the model response is evaluated for a series of cracks that are relatively close in terms of size, orientation, and location. Accordingly, TL is expected to provide advantages for cracks generated through mutation, but not through crossover.

\begin{figure}[!ht]
        \centering
        \includegraphics[width=0.6\textwidth]{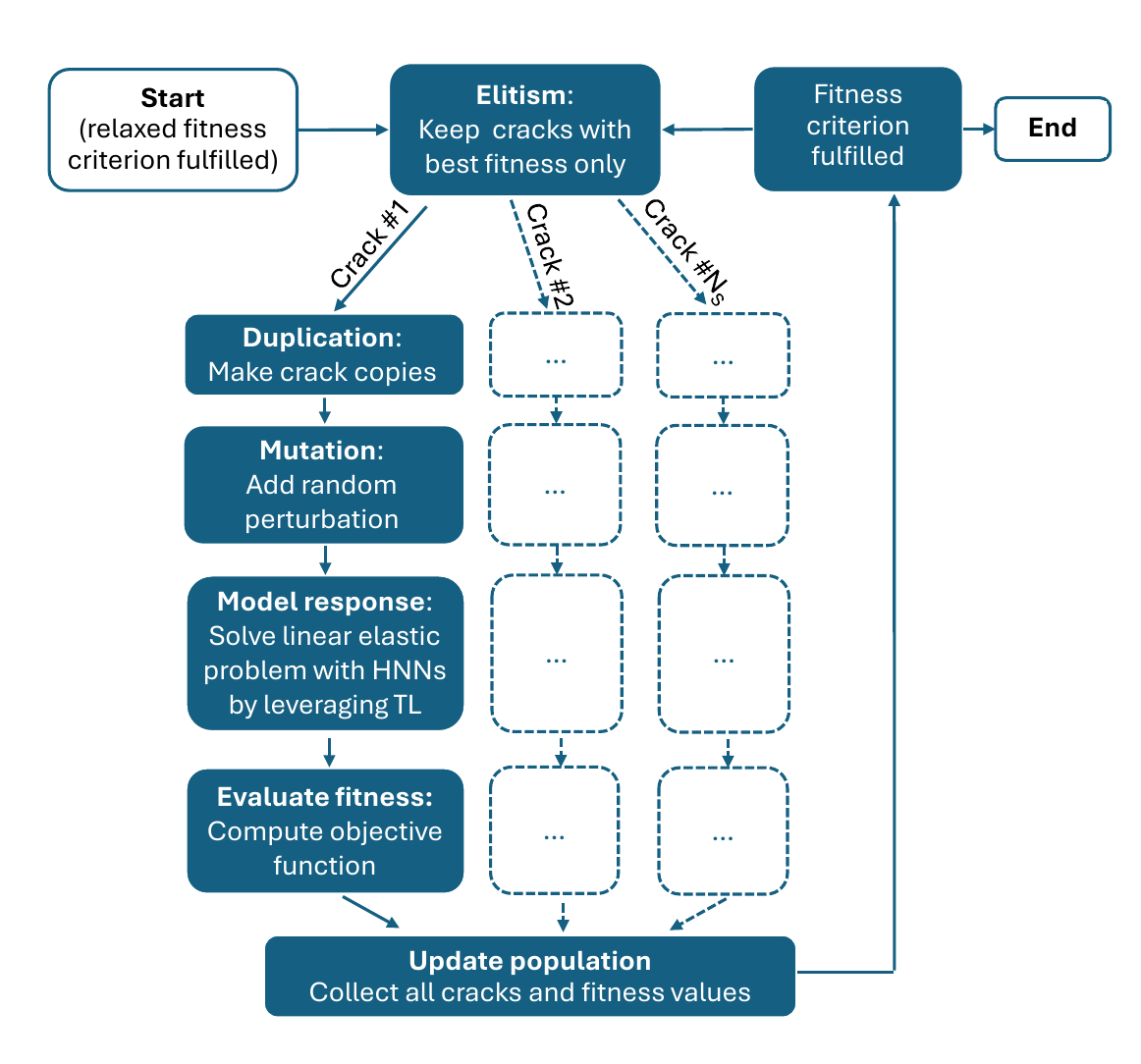}
        \caption{New algorithm for short-range search, which exploits HNNs and TL to efficiently evaluate the model response.}
        \label{fig:NJT_algorithm}
\end{figure}

Based on these considerations, a new crack detection strategy is proposed, where the long-range and short-range searches are decoupled. Specifically, the algorithm shown in \Cref{fig:waisman_algorithm} is retained for the long-range search, but with two important modifications. First, the model response is evaluated using the HNN-based approach described in \Cref{sec:HNNs}, instead of XFEM. Second, the algorithm terminates once $N_S$ cracks are identified that satisfy 
\begin{equation}\label{eq:fitness_switch}
    \mathcal{F}(\bm{p}) < \mathcal{F}_{s}
\end{equation}
where $\mathcal{F}_{s}>\mathcal{F}_{t}$ is a relaxed fitness threshold. After termination, the algorithm shown in \Cref{fig:NJT_algorithm} is initiated, which refines the solution by exploring the search space in the vicinity of the cracks identified in the previous stage. For each crack provided as input, this second algorithm creates $N_D$ duplicates, which undergo mutation with unit probability according to \Cref{eq:mutation}. Subsequently, the model response is evaluated for the mutated duplicates using the HNN-based approach combined with TL. As the mutated duplicates do not differ substantially from each other, ideal conditions for TL exist. All input cracks and their mutated counterparts are then collected to form a new population. If this population contains at least one crack that satisfies the fitness criterion \labelcref{eq:fitness_threshold}, the algorithm terminates. Otherwise, the $N_S$ cracks with the best fitness are retained, while the others are discarded, and a new iteration is performed.

It is worth noting that, in addition to enabling TL, the use of HNNs in place of XFEM offers an additional advantage. Specifically, it allows control over the number of training epochs used to compute the model response. This parameter has no direct counterpart in XFEM. In fact, the primary parameter controlling the solution accuracy in XFEM is the mesh size, which reflects the capability of the underlying numerical approximation to represent functions of a given complexity. In HNNs, this role is played by the network architecture – namely, the number of layers and neurons. 

While a large number of epochs may be required to achieve the maximum accuracy permitted by the underlying network representation, such a large number may not be necessary in the present context. In fact, a smaller number may yield an estimate of the model response that is less accurate, but still of sufficient quality within the framework of the algorithms shown in \Cref{fig:waisman_algorithm,fig:NJT_algorithm}. This aspect can have a beneficial impact on the algorithm computational efficiency, which will be investigated in \Cref{sec:convergence}.

\subsection{Numerical experiments}\label{sec:num_experiments}
\begin{figure}[!ht]
    \centering
    \begin{subfigure}[b]{0.32\textwidth}
         \centering
         \includegraphics[width=\textwidth]{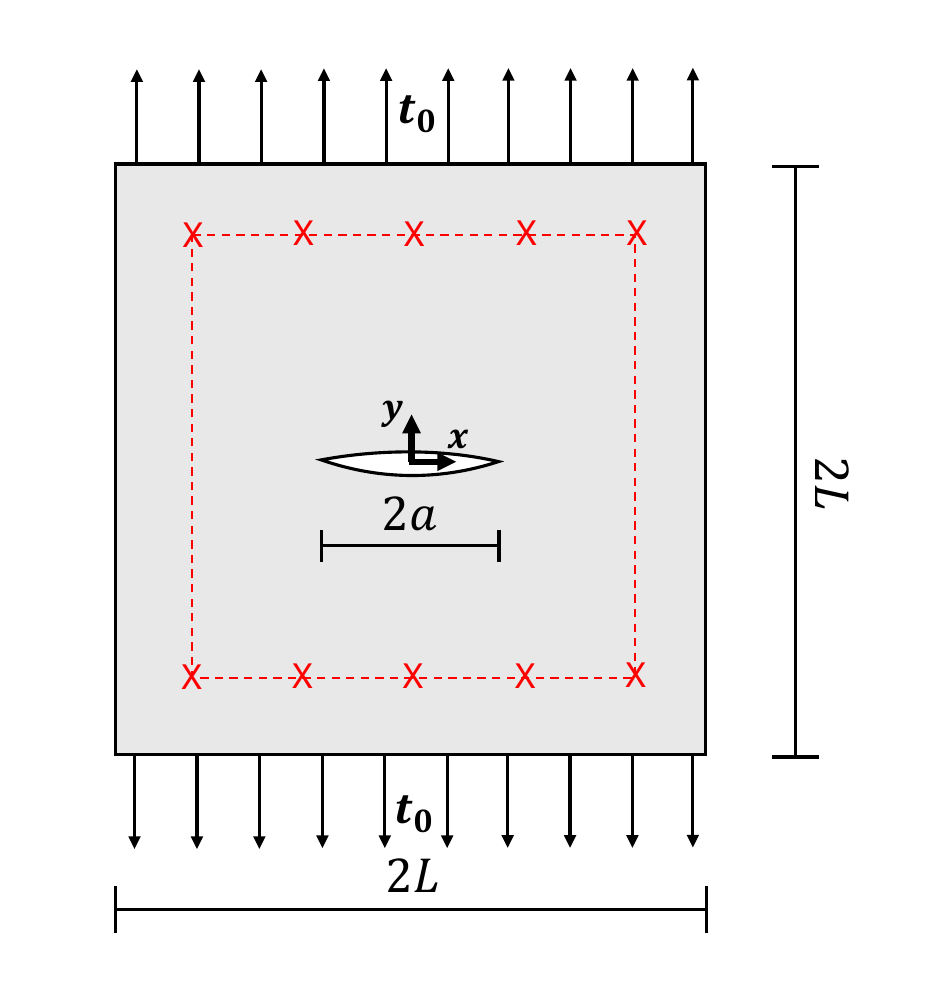}
         \caption{}
         \label{fig:num_exp_1}
     \end{subfigure}
     \begin{subfigure}[b]{0.32\textwidth}
         \centering
         \includegraphics[width=\textwidth]{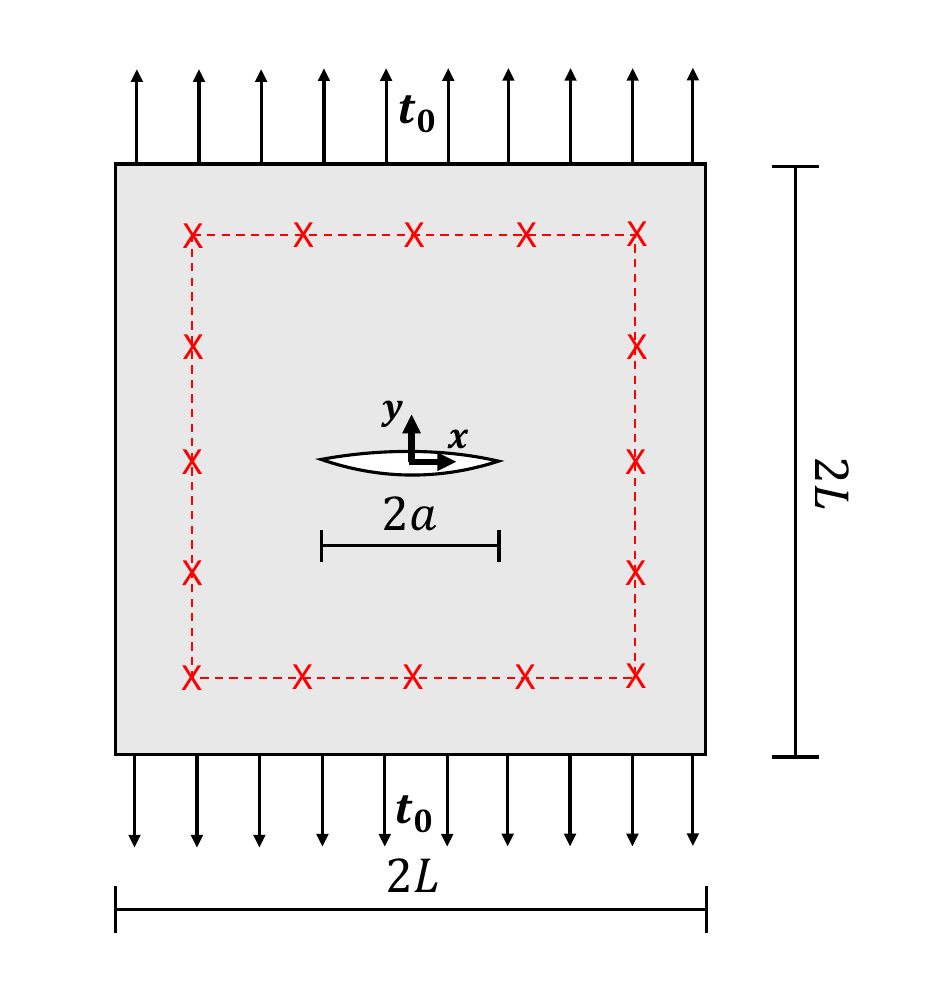}
         \caption{}
         \label{fig:num_exp_2}
     \end{subfigure}
     \begin{subfigure}[b]{0.32\textwidth}
         \centering
         \includegraphics[width=\textwidth]{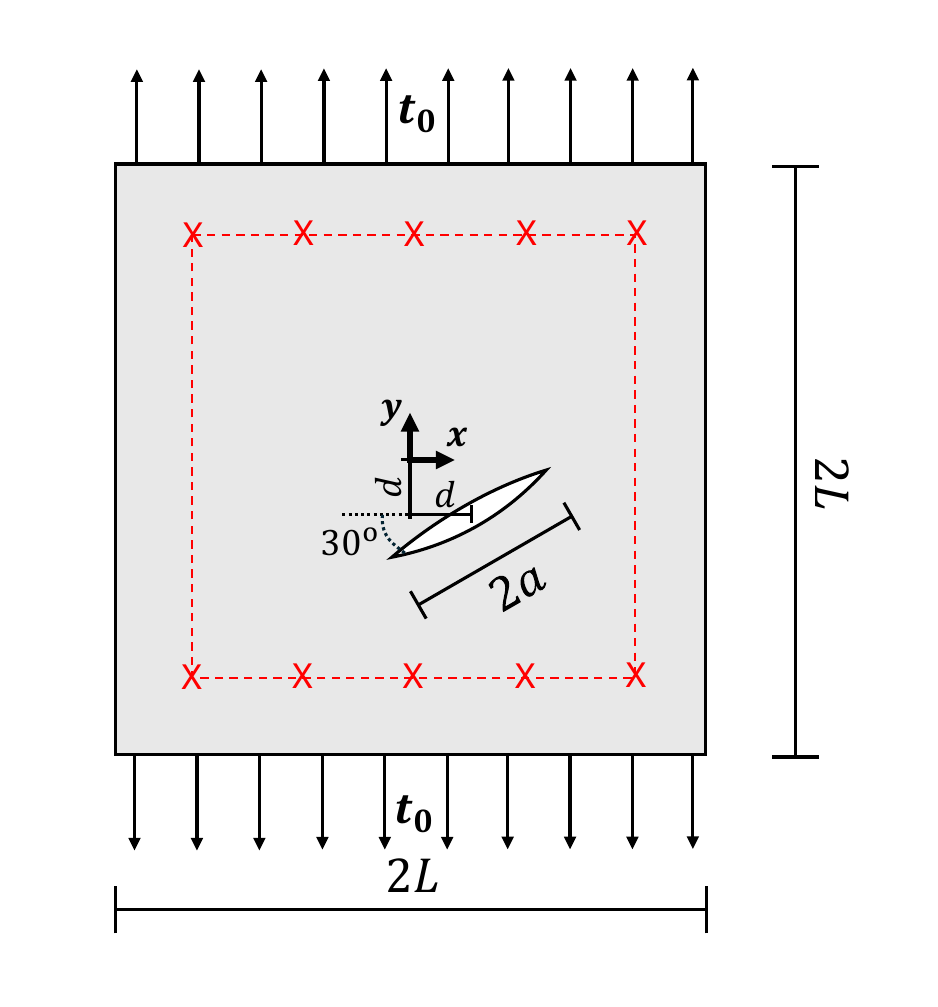}
         \caption{}
         \label{fig:num_exp_3}
     \end{subfigure}
    \caption{Geometry, boundary conditions, sensor placement (indicated by the red crosses), and search space (indicated by the red dashed box) for the numerical experiments. (a) Plate with central crack subjected to  uniform tensile loading. (b) Same as in (a), but with 16 strain sensors instead of 10. (c) Same as in (a), but with slanted crack.}
    \label{fig:num_exp}
\end{figure}

Three numerical experiments are conducted to assess the crack detection strategy based on GA–HNNs proposed in \Cref{sec:GA-HNNs} and to compare it with the reference GA–XFEM approach of \Cref{sec:GA-XFEM}. The first experiment involves detecting a central crack in a square plate subjected to uniform tension on the top and bottom edges, see \Cref{fig:num_exp_1}. The ratio of crack length to plate side length is $ a/L = 3/10 $ and the ratio of the elastic parameters is $\lambda/\mu = 1$. The magnitude of the applied traction vector is $\Vert \bm{t_0}\Vert=\lambda $ and plane strain conditions are assumed. Strain measurements are obtained from 10 sensors, placed symmetrically at $y = \pm 0.8L$ and $x = \{-0.8L;-0.4L;0;0.4L;0.8L\}$ with respect to the global coordinate system positioned at the center of the plate. 

The second crack detection experiment, shown in \Cref{fig:num_exp_2}, differs from the first only in terms of the number of sensors. In fact, 6 additional sensors are placed symmetrically at $x = \pm 0.8L$ and $y = \{-0.4L;0;0.4L\}$. This experiment is meant to provide indications on the sensitivity of the detection algorithm to the sensor density. Finally, the third experiment has the same sensor placement as the first, but different crack configuration, see \Cref{fig:num_exp_3}. Specifically, the crack is rotated anticlockwise by $ 30 ^{\circ}$ with respect to the x-axis, and its center is located at $ x=+d$ and $y=-d$, with $d=L/5$. 

For all three numerical experiments, the strain measured by the sensors is obtained using a high-fidelity finite element analysis performed in Abaqus. The analysis employs a crack-conforming mesh composed of second-order quadrilateral elements, with an element size ranging from $0.01L$ at the plate boundary to $0.002L$ in the vicinity of the crack tips. A sweep meshing strategy is adopted near both crack tips, and collapsed quarter-point elements are used to accurately capture the elastic singularities.

\section{Results}\label{sec:results}
The GA–HNN and the GA–XFEM crack detection strategies were both implemented in Python\footnote{\url{https://github.com/jonashund/holomorphic_crack_detection}}. To evaluate the model response, the GA-HNN approach leverages the original PIHNN library \cite{calafa2024}, while the GA–XFEM approach employs Abaqus XFEM. 

\subsection{Convergence analysis}\label{sec:convergence}
As mentioned in \Cref{sec:GA-HNNs}, the number of training epochs is expected to have a significant impact on the convergence behavior – and thus the efficiency – of the proposed GA–HNN crack detection strategy. To investigate this aspect, the first numerical experiment shown in \Cref{fig:num_exp_1} is considered. The optimization problem is solved using the parameter values listed in \Cref{tab:GA_param}. The search space is defined as $\mathcal{P} = [-0.8L, +0.8L]\times[-0.8L, +0.8L]$, and a minimum admissible crack length of $0.1L$ is imposed. The model response is evaluated by training two holomorphic neural networks, each composed of three fully connected hidden layers with ten neurons per layer. The values of the main hyperparameters used for the training are reported in \Cref{tab:GA_param}. A total of $N_\sigma = 300$ training points are employed, which are uniformly distributed along the plate boundary. Given the relative simplicity of the boundary geometry and boundary conditions, this value of $N_\sigma$ is sufficient to ensure an accurate computation of the model response, provided that the number of training epochs is large enough.

\begin{table}[h!]
    \centering
    \begin{tabular}{ |c c | c c | c c | c c c |}
         \hline
         \multicolumn{2}{|c|}{Initialization} & \multicolumn{2}{|c|}{Crossover} & \multicolumn{2}{|c|}{Mutation} & \multicolumn{3}{|c|}{Short-range search}\\
         \hline
         $N_I$ & $ T $ & $N_{\Delta}$ & $N_O/N_P$ & $ P_M$ & $\delta_m/L$ & $ \mathcal{F}_s $ & $N_S $ & $N_D $ \\
         \hline
         9     & 5     & 1            & 1         &  0.5   &     0.1     &   0.007           & 3      & 2 \\ 
         \hline
    \end{tabular}
    \caption{Values of the genetic optimization parameters adopted in this work. Values related to initialization, crossover and mutation are from \citet{waisman_detection_2010}.}
    \label{tab:GA_param}
\end{table}

\begin{table}[h!]
    \centering
    \begin{tabular}{ |c c  c c c|}
         \hline
         Optimizer & Learning rate & $s_u$ & $s_\sigma$ & $N_\sigma $\\
         \hline
         ADAM & $10^{-2}$  & $1$ & $1$ & $300 $\\
         \hline
    \end{tabular}
    \caption{Main hyperparameters used to train the holomorphic neural networks.}
    \label{tab:hyperparam}
\end{table}

\begin{figure}[!ht]
        \centering
        \includegraphics[width=0.67\textwidth]{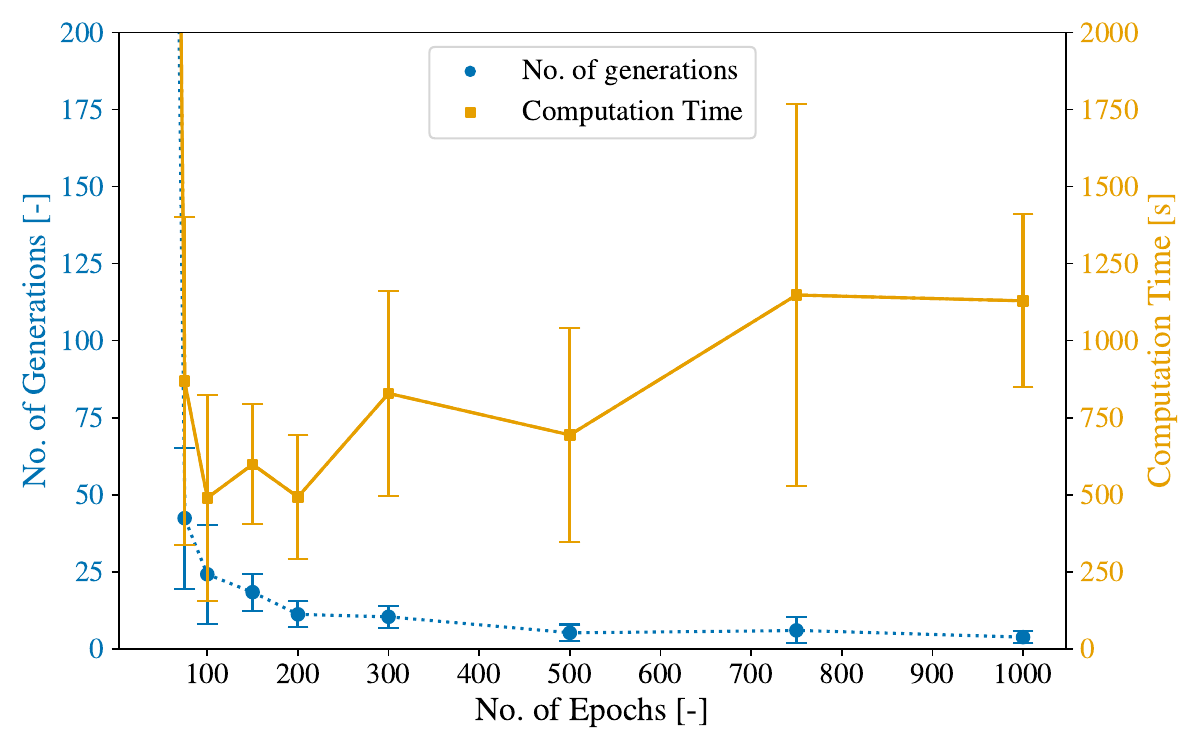}
        \caption{Impact of the number of epochs on the number of generations and computation time for the long-range search of the GA-HNN crack detection strategy. The problem of \Cref{fig:num_exp_1} is considered, and the markers indicate the mean of five runs per epoch with identical initial population.}
        \label{fig:impact_n_epochs}
\end{figure}

\Cref{fig:impact_n_epochs} shows the CPU time and the number of generations required to complete the long-range search as a function of the number of training epochs. It can be observed that when the number of epochs drops below 100, both CPU time and the required number of generations increase dramatically. This occurs because the number of epochs is insufficient, leading to large errors in the computed model response, and consequently the crack detection algorithm struggles to converge. When the number of epochs is reduced below 50, no convergence is achieved at all. Even if the “correct” crack configuration is generated during the optimization, the algorithm may fail to recognize it because its fitness is evaluated inaccurately. The same figure also shows that increasing the number of epochs decreases the number of generations monotonically, until a plateau is reached. This behavior is expected: the plateau corresponds to the minimum number of epochs required to accurately capture the model response. Once this threshold is reached, further increasing the number of epochs only increases CPU time, with minimal gains in accuracy. Therefore, an optimal number of epochs exists that balances accuracy and computational cost. For the present problem, a value of approximately 200 epochs provides this balance and is adopted in the remainder of the article.

\begin{figure}[!ht] 
    \centering
    \includegraphics[width=0.66\linewidth]{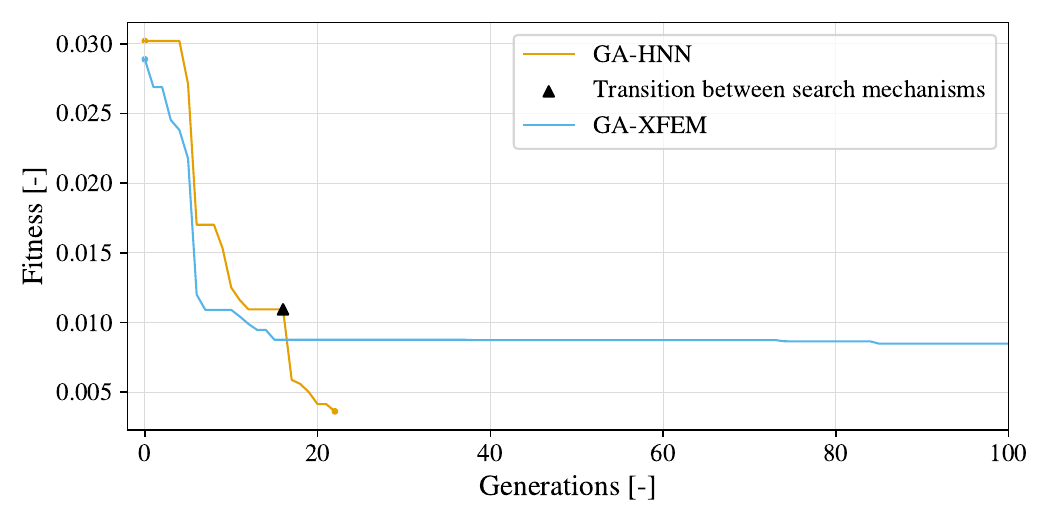}
    \caption{Evolution of the population fitness as a function of the number of generations. The curves shown are for individual runs of the GA-HNN and GA-XFEM crack detection strategies applied to the problem of \Cref{fig:num_exp_1}. The runs are selected to highlight the typical convergence behavior of GA-HNN, and the slow rate of solution refinement of GA-XFEM, respectively.}
    \label{fig:fitness_vs_n_gen}
\end{figure}

Additional insight into the convergence behavior of the proposed GA–HNN approach is provided by \Cref{fig:fitness_vs_n_gen}, which shows the evolution of the population fitness – defined as the minimum value of the objective function in \Cref{eq:obj_function} across all cracks – as a function of the number of generations. Because of the inherent randomness of the genetic optimization procedure, each run of the algorithm produces a different fitness curve. The curve shown in the figure is therefore selected as representative of the behavior observed over multiple runs. The plot reveals that the optimization progresses in a stepwise manner, alternating periods in which the fitness remains constant with sudden drops corresponding to the discovery of a new crack with improved fitness. The transition from the long-range to the short-range search is indicated by the black triangle and is followed by a rapid decrease in the objective function, which arises because the search becomes restricted to smaller regions around the best cracks in the population. In genetic terms, the transition corresponds to the algorithm shifting its emphasis from generating markedly different offspring through crossover to refining existing individuals through mutation. 

For comparison, a curve corresponding to a single run of the GA–XFEM algorithm is also shown in \Cref{fig:fitness_vs_n_gen}. The curve indicates that although convergence is initially rapid, it stagnates after approximately 15 generations. This behavior is not uncommon for the GA–XFEM approach \cite{waisman_detection_2010} and can be attributed to the fact that, if the search domain is not progressively restricted, standard genetic algorithms may quickly identify a good approximation of the solution, but the subsequent refinement proceeds slowly. This limitation becomes particularly detrimental when model evaluations are computationally expensive, as is typically the case in crack detection problems. Therefore, it can be concluded that introducing a short-range search step is beneficial, independently of the fact that, in the present work, it is also used to enable TL. This observation is consistent with previous findings which, in the more general context of multiple-crack detection, motivated the development of multiscale detection strategies \cite{sun_multiscale_2014,zhao_adaptive_2018}.

\begin{figure}[!ht]
    \centering
    \includegraphics[width=0.9\linewidth]{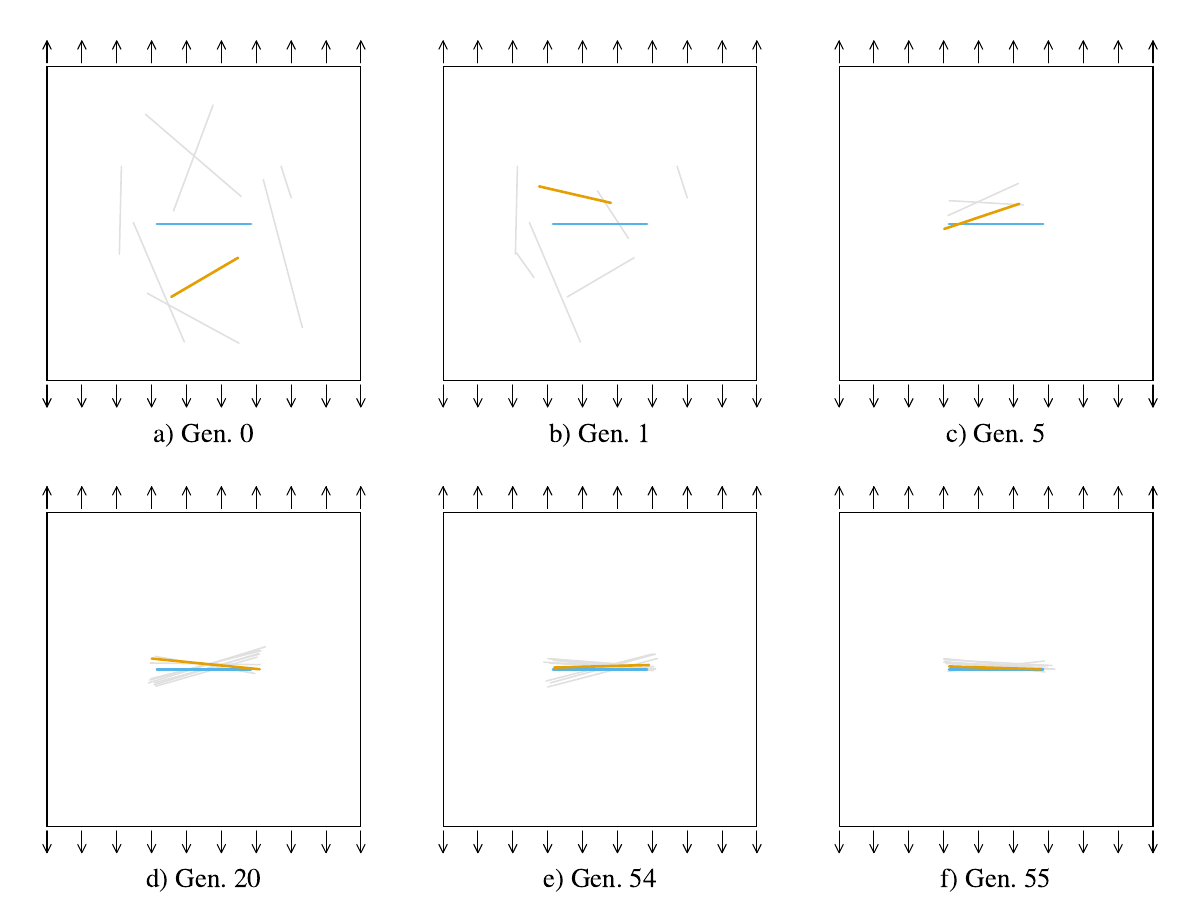}
    \caption{Snapshots of the crack population at selected generations, for a single run of the GA-HNN crack detection strategy applied to the problem of \Cref{fig:num_exp_1}. (a) The initial population, (b) Generation 1, (c) Generation 5, (d) Generation 20, (e) Generation 54, and (f) The final population in generation 55. The target crack is in light blue, the current fittest crack in yellow, and the rest of the population in gray.}
    \label{fig:pop_snapshots}
\end{figure}

While \Cref{fig:fitness_vs_n_gen} illustrates the evolution of the fittest crack in the population, it does not provide insight into how the entire population evolves over successive generations. This aspect is clarified in \Cref{fig:pop_snapshots}, which shows snapshots of the crack population at increasing generation numbers for a representative run of the GA–HNN algorithm. The first snapshot exhibits a wide dispersion in crack size, orientation, and location, consistent with the random initialization. The next two snapshots, corresponding to generations 1 and 5, show how this dispersion decreases as cracks with low fitness are progressively eliminated. These snapshots also reveal the reduction in population size that precedes the periodic, partial re-initialization step discussed in \Cref{sec:GA-XFEM}, which is triggered in generation 6. The final three snapshots, corresponding to generations 20, 54, and 55, are taken after the algorithm enters the short-range search. At this stage, the cracks become highly clustered, and the clusters progressively narrow until they converge toward the correct solution.

The above results show that the short-range search has a significant impact on convergence, and thus the optimal selection of the associated search parameters has to be discussed. While increasing the values of $N_s$ and $N_D$ has only a limited influence on the behavior of the algorithm, varying the parameter $\mathcal{F}_s$ – which governs the transition between the two search mechanisms – has a pronounced effect. If $\mathcal{F}_s$ is set to a very large value, crack identification terminates before the short-range search is activated. Conversely, if $\mathcal{F}_s$ is set to a very small value, the short-range search begins immediately, meaning that the optimization proceeds in the neighborhood of cracks that have been generated randomly rather than selected on the basis of their fitness. In both cases, convergence is slowed down, highlighting the importance of an appropriate choice of $\mathcal{F}_s$. The optimal value of $\mathcal{F}_s$, however, is problem dependent to some extent, as we verified through tests involving different sensor numbers and locations. A pragmatic way to determine a suitable value is to monitor the evolution of the population and set $\mathcal{F}_s$ such that the short-range search is triggered once a few high-fitness cracks have emerged and persisted across generations. Although this method is applied manually in the present work, it could be automated in future studies using machine learning techniques.

\subsection{Efficiency comparison}\label{sec:efficiency}
A primary objective of the present work is to assess whether the proposed GA–HNN crack detection strategy can be competitive, in terms of computational efficiency, with the well-established GA–XFEM approach. To this end, both strategies are applied to the three numerical experiments shown in \Cref{fig:num_exp}. Since the two strategies differ markedly – particularly in the way the model response is evaluated – special care is taken to ensure the fairest possible comparison, as detailed in the following.

First, parameters common to both the GA–XFEM and GA–HNN algorithms are assigned the same values. Specifically, since the long-range search in GA–HNN is conceptually identical to the search in GA–XFEM – apart from the way the model response is evaluated – the crossover and mutation parameters are set to the values reported in \Cref{tab:GA_param}. Regarding initialization, the same randomly generated population is supplied to both algorithms to ensure identical conditions at generation zero. Furthermore, although both crack detection strategies are suitable for parallelization, all computations are carried out on a single CPU. This choice is made for simplicity, as a detailed analysis of the scalability of the two approaches is beyond the scope of the present work. It is also worth noting that, unlike XFEM, HNNs typically perform best on GPUs, implying that the results presented here are expected to be conservative with respect to the computational performance of the GA–HNN algorithm.

Parameters specific to the GA–HNN strategy, including the hyperparameters controlling the training of the HNNs, are set as described in \Cref{sec:convergence}. Regarding XFEM, a regular rectangular mesh and the default Abaqus selection of all the XFEM parameters are adopted. The crack is modeled as a wire feature, and the feature end points are updated each time the model response is to be evaluated, without modifying the underlying mesh.

The comparison between the two crack detection strategies is performed under the conditions that (1) the model response is represented with the same level of accuracy, and (2) the final identified crack achieves the same level of accuracy with respect to the reference solution. Regarding point (1), \Cref{sec:GA-HNNs} remarked that the representation accuracy is governed by the network architecture in the HNN approach and by the mesh size in the XFEM approach. As demonstrated in \cite{calafa_solving_2025} for problems closely related to those shown in \Cref{fig:num_exp}, using HNNs with 3 hidden layers of 10 neurons each yields stress and strain fields with an accuracy comparable to that obtained using an XFEM mesh size of $0.06L$. Accordingly, the same settings are adopted here. It is worth noting that the chosen XFEM mesh size is of the same order as that used by \citet{waisman_detection_2010} for similar benchmark configurations. Regarding point (2), comparable solution accuracy is ensured by introducing a special termination criterion: both the GA-XFEM and GA-HNN algorithms are terminated once a crack is found whose tips lie within a radius of $0.05a$ from the corresponding tips of the target crack.

\begin{figure}[ht]
    \centering
    \includegraphics[width=0.8\linewidth]{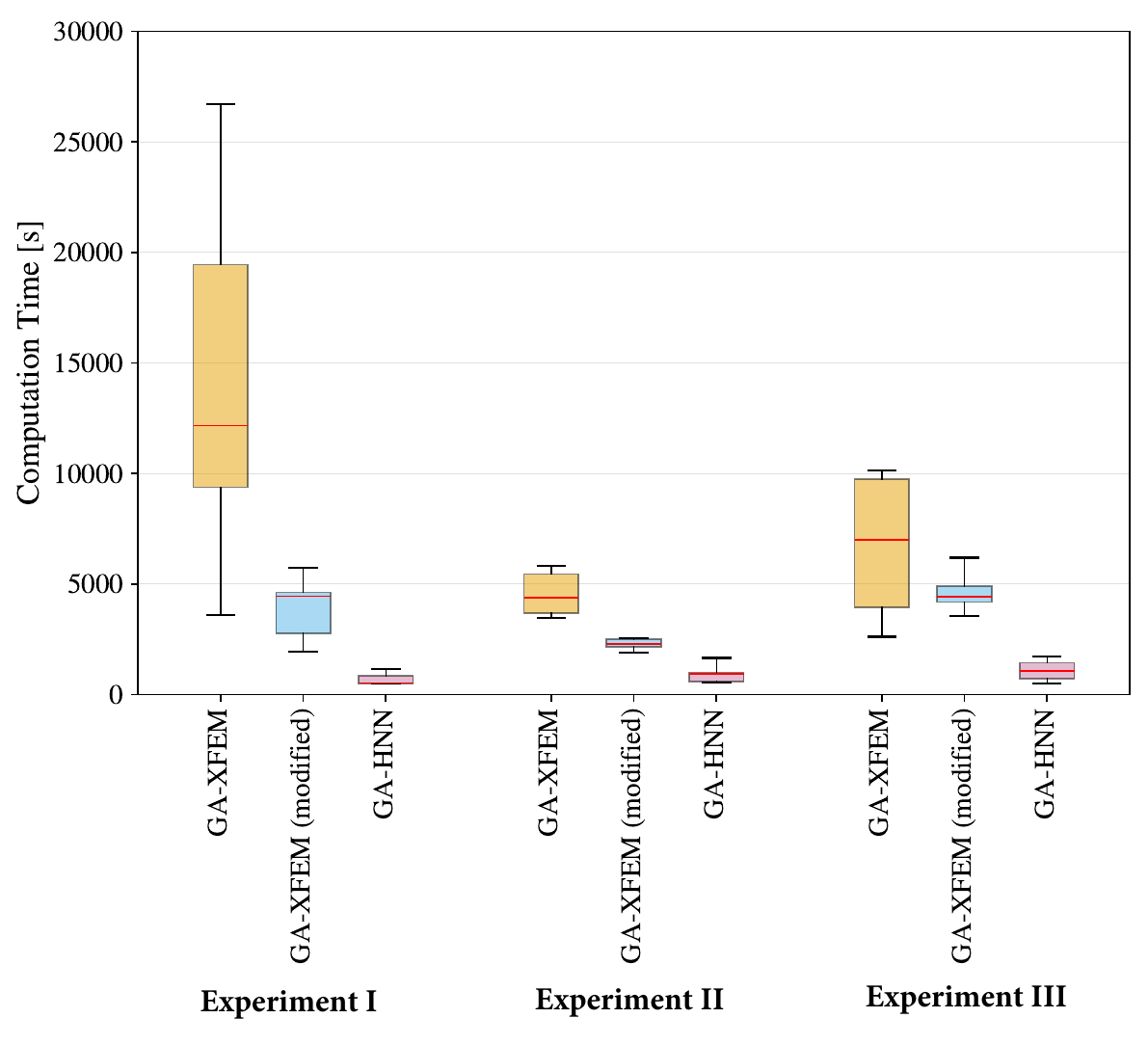}
    \caption{Efficiency of the crack detection strategies in terms of CPU time. Each box is based on five independent runs. The box represents the interquartile range (Q1-Q3), the whiskers indicate the full range of the data, and the horizontal line denotes the median.}
    \label{fig:box_plot_computation_time}
\end{figure}

\Cref{fig:box_plot_computation_time} shows the total CPU time required by the two crack detection strategies when applied to the three numerical experiments. The results are presented as box plots, where each box corresponds to five independent runs of the corresponding algorithm. In addition to the GA–XFEM and GA-HNN approaches, a third variant is considered, denoted as “GA–XFEM (modified)”, which consists of running the GA–XFEM strategy using the same two-step search mechanism employed in the GA–HNN approach. This modification enables a separate assessment of the impact of introducing the short-range search within the GA loop and of the effect of replacing XFEM with HNNs for evaluating the model response.

The figure shows that the median CPU time of the GA–HNN approach is of the same order across the three experiments, ranging from 527 s for experiment I to 1339 s for experiment III. The median CPU time of the GA–XFEM approach is consistently larger, by a factor ranging from 23 in experiment I to 7 in experiment III. Moreover, its variation across experiments is noticeably greater. In particular, the GA–XFEM approach appears more sensitive to the number of strain sensors, as evidenced by the substantial spread observed between experiments I and II. This pronounced variability is consistent with previous findings that highlighted the strong sensitivity of GA–XFEM to sensor number and placement \cite{waisman_detection_2010}. 

The modified GA–XFEM approach exhibits median CPU times that systematically lie between those of the two other strategies. This indicates that the reduction in CPU time achieved by the GA–HNN approach stems from two distinct sources. The first is the more efficient search mechanism, which employs separate long-range and short-range phases executed sequentially. The second is the more efficient evaluation of the model response, particularly during the short-range search where TL is exploited.

\begin{figure}[ht]
    \centering
    \includegraphics[width=0.8\linewidth]{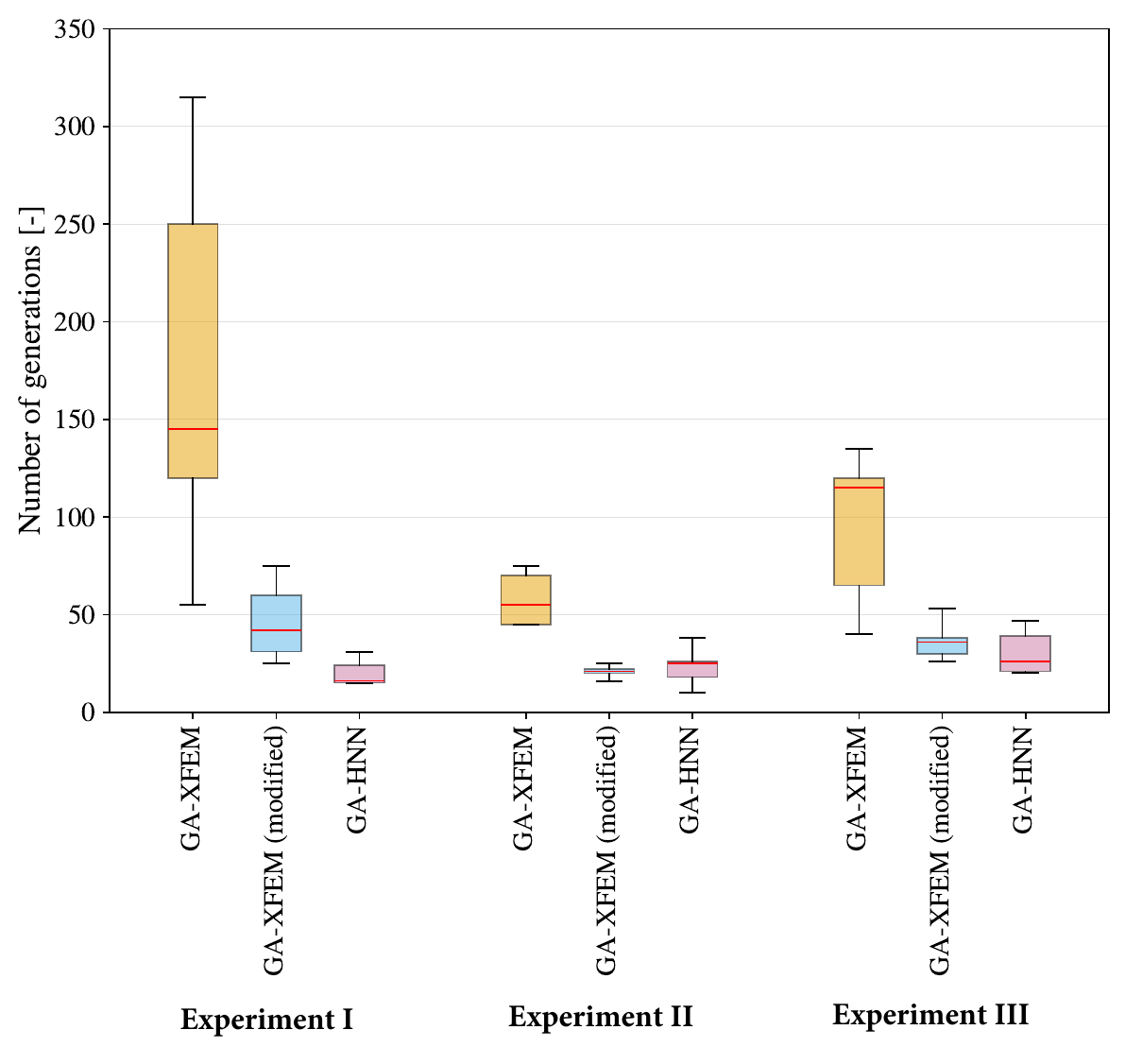}
    \caption{Efficiency of the crack detection strategies in terms of number of generations. Each box is based on five independent runs. The box represents the interquartile range (Q1-Q3), the whiskers indicate the full range of the data, and the horizontal line denotes the median.}
    \label{fig:box_plot_num_generations}
\end{figure}

The increased efficiency achieved by combining HNNs with transfer learning compared to XFEM is further supported by the results in \Cref{fig:box_plot_num_generations}, which reports the number of generations required by the three approaches for all numerical experiments. It can be seen that for experiment I, the modified GA–XFEM approach requires more generations than the GA–HNN approach. However, for experiments II and III, both approaches require the same number of generations, indicating that the corresponding difference in CPU time must be attributed to the faster evaluation of the model response provided by HNNs. The figure also confirms the improved efficiency of the two-stage search mechanism: the median number of generations for the reference GA–XFEM approach is consistently larger than that observed for the other two approaches.

\subsection{Noise sensitivity}\label{sec:noise_sensitivity}
The results presented in the previous sections are based on the assumption that the strain values measured by the sensors are error free. However, in real-world scenarios, strain data are inevitably corrupted by noise to some extent. It is therefore important to analyze how the proposed GA–HNN crack detection strategy is affected by the presence of noise in the strain measurements.

To this end, crack identification is repeated for the second numerical experiment shown in \Cref{fig:num_exp_2}, assuming that the strain values measured by the 16 sensors are corrupted by zero-mean Gaussian white noise. Specifically, the strain component $\upvarepsilon_{ij}$ measured by the n-th sensor is assumed to be given by:
\begin{equation}\label{eq:noise_formula}
    (\upvarepsilon_{ij})_n = (\upvarepsilon_{ij})_n^{ref} + Ym \sqrt{\frac{1}{n_s}\sum_{n=1}^{n_s} \left( (\upvarepsilon_{ij})_n^{ref} \right)^2}
\end{equation}
In the above expression, the first term on the right-hand side represents the reference, noise-free strain. The second term corresponds to the root-mean-square value of the noise-free strain taken over all sensors, multiplied by a normally distributed random variable $Y$ with zero mean and unit variance, and by a real-valued parameter $m$ that controls the noise level.
Both the GA–HNN and GA–XFEM algorithms are run on the noisy strain data for a fixed number of 50 generations. The resulting level of accuracy is then quantified by computing the crack detection error, defined as:
\begin{equation}\label{eq:crack_det_error}
    E = |z^+_T - z^+_F| + |z^-_T -  z^-_F| 
\end{equation}
where $z^+_T, z^-_T$ are the locations of the right and left tips of the true crack, and $z^+_F, z^-_F$ are the corresponding tip locations for the crack in the population that exhibits the best fitness. The procedure is repeated for values of $m$ equal to 0\%, 5\%, 10\% and 15\%, and for each $m$ value, 10 independent runs of the algorithms are considered to get statistical insight.

\begin{figure}[ht]
    \centering
    \includegraphics[width=0.67\textwidth]{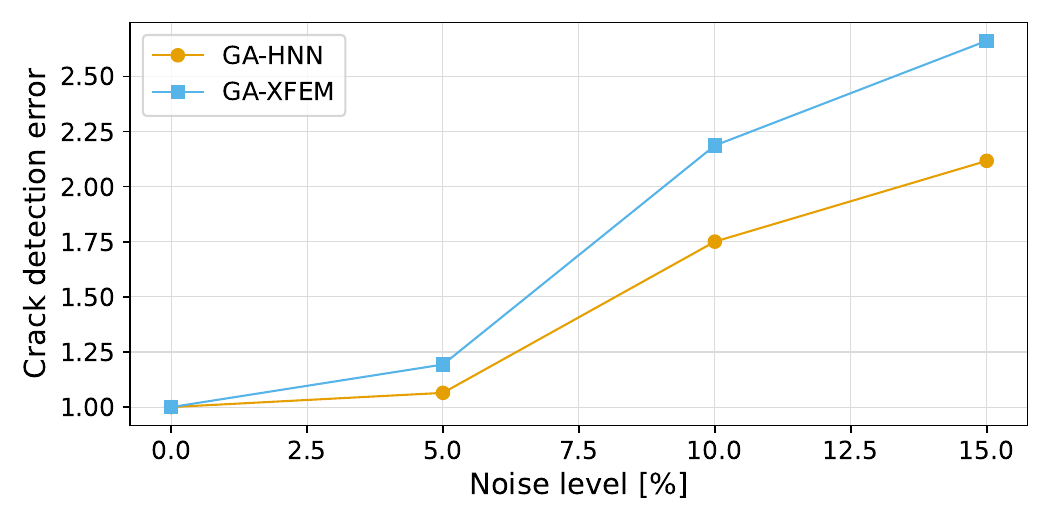}
    \caption{Impact of strain measurement noise on the accuracy of the GA–HNN and GA–XFEM crack detection strategies. For each noise level, the accuracy is expressed in terms of the crack detection error defined in \Cref{eq:crack_det_error}. The reported values correspond to the mean over ten independent runs and are normalized by the mean error obtained in the noise-free case.}
    \label{fig:noise_sensitivity}
\end{figure}

\Cref{fig:noise_sensitivity} presents the evolution of the crack detection error as a function of the noise level. For both the GA–HNN and GA–XFEM strategies, the data are normalized by the error obtained in the noise-free case, in order to highlight the relative effect of noise. As expected, the crack detection error increases monotonically with increasing noise level, since higher noise renders the inverse problem less well posed. For a noise level of 15\%, the error is approximately 210\% larger than in the noise-free case for the GA–HNN approach, and about 270\% larger for the GA–XFEM approach. In this respect, it is worth noting that the GA–HNN strategy is less affected by noise for all noise levels considered. This consistent behavior suggests that the GA–HNN approach may be intrinsically more robust to noise than GA–XFEM. This improved robustness may be attributed to the different manner in which the model response is evaluated, with GA–HNN relying on an approximation of the response, as discussed in \Cref{sec:convergence}.   

\section{Conclusions}\label{sec:conclusions}
A physics-informed machine learning framework based on holomorphic neural networks is introduced for detecting cracks in two-dimensional solids from strain or displacement data. Crack detection is formulated as an inverse problem and solved using genetic optimization. At each generation, the fitness function is evaluated by expressing the solution to the corresponding elasticity problem in terms of two holomorphic potentials, which are subsequently found by training two HNNs. As the potentials automatically satisfy equilibrium and traction-free conditions along the crack faces, the training proceeds quickly based solely on the boundary conditions. Training efficiency is further improved by splitting the genetic algorithm into long-range and short-range stages, enabling the use of transfer learning in the latter.

The new strategy has been assessed through three benchmark tests. The results show that an optimal number of epochs exists for training the HNNs, which provides a balance between accuracy and computational cost and thereby yields the best overall performance. The proposed strategy has also been compared with the reference approach, in which XFEM is used to compute the model response. Using the requirement of identical stress-field representation accuracy to determine both the HNN architecture and the XFEM mesh size, the proposed method is found to be between 7 and 23 times faster than the standard XFEM-based approach. This improvement arises partly from the more efficient two-stage search mechanism, and partly from the more rapid evaluation of the model response enabled by transfer learning. Furthermore, the proposed strategy appears to exhibit greater robustness to strain measurement noise than the standard XFEM-based approach.

The present study builds on strong assumptions, as it considers the simplified case of a single internal crack. Extending the approach to an edge crack would be straightforward, requiring only an update of the expressions for the complex potentials in \Cref{eq:rice_mod} \cite{calafa_solving_2025}. Extension to multiple cracks would be more challenging instead. While domain decomposition techniques \cite{calafa2024} can, in principle, enable the HNN approach to compute the elastic response of a body containing multiple cracks, such a decomposition requires a priori knowledge of the total number and approximate locations of the cracks, which are typically unknown in multi-crack detection problems. Nevertheless, approaches based on multiscale algorithms \cite{sun_multiscale_2014} and the introduction of topological variables \cite{sun_nondestructive_2013} may provide viable strategies to overcome this limitation.

In the light of the above, the present findings demonstrate that combining genetic optimization with HNNs and transfer learning offers a promising avenue for developing crack detection strategies with higher efficiency than those currently available.

\section*{CRediT authorship contribution statement}
\textbf{Jonas Hund}: Software, Formal analysis, Visualization. 
\textbf{Nicolas Cuenca}: Conceptualization, Software, Formal analysis. 
\textbf{Tito Andriollo}: Conceptualization, Writing – original draft, Supervision, Funding acquisition.

\section*{Declaration of competing interest}
The authors declare that they have no known competing financial interests or personal relationships that could have appeared to influence the work reported in this paper.

\section*{Acknowledgments}
This work was supported by the Aarhus University Research Foundation, Denmark through the grant no. AUFF-E-2023-9-44 “Strength of materials-informed neural networks”.

\section*{Data availability}
The Python code associated with the present work is available at \url{https://github.com/jonashund/holomorphic_crack_detection}.

\normalsize
\bibliography{references}

\end{document}